\documentclass[12pt,fleqn]{article}
\usepackage{amssymb,graphicx,graphics,epsfig}
\usepackage{lscape,graphics,amsmath}
\usepackage{latexsym,amsfonts}
\usepackage{xcolor}
\usepackage{ulem}
\usepackage{cite}
\begin{document}

\begin{center}
{\Large  {\bf On Compton ionization of a hydrogen atom by twisted photons} \\
\vspace{1cm} I.P.~Volobuev, V.O.~Egorov}

\vspace{5mm}

{Skobeltsyn Institute of Nuclear Physics, Lomonosov Moscow
State University,\\ Moscow 119991, Russia}
\end{center}
\vspace{0.5cm}

\begin{abstract}
The differential probability of the process of Compton ionization
of a hydrogen atom by a cylindrical electromagnetic wave is
calculated taking into account the finite size of the target,
which resulted in the appearance of a dependence of this value on
the angular momentum of the cylindrical wave. It is shown that the
use of a cylindrical wave instead of a plane wave does not lead to
new angular distributions of the reaction products.
\end{abstract}

\section{Introduction}

The interaction of particles in twisted quantum states with atomic
targets has  been widely discussed for the last 25 years or so,
and a large number of papers have been published on this topic
\cite{Molina-Terriza,Harris,arxiv,TwAt,Serbo,IS,Ivanov,PRL,Kiselev}.
Twisted electrons are discussed, for example, in \cite{Harris,
arxiv}, and  paper \cite{arxiv} is of a review nature. There are
papers on twisted neutrons and even atoms \cite{TwAt}, but most of
the papers are devoted to twisted photons
\cite{Serbo,IS,Ivanov,PRL,Kiselev}. The cited papers deal with
various aspects of twisted photon scattering on atoms, including
photoionization, inverse Compton scattering, photoexcitation, and
the like. It should be noted that this issue is most fully
presented in  recent review \cite{Serbo}, which also includes an
extensive list of literature.

Usually twisted photons are considered, which are described by a
Bessel cylindrical wave, although more complex wave configurations
are also discussed \cite{NJP}. It should be noted right away that,
unlike a plane wave, the field of a cylindrical wave is not
translationally invariant in the plane orthogonal to the direction
of its propagation, so the result of a scattering process should
depend both on the position of the scatterer, which in this case
is characterized by an impact parameter, and on its size.
Therefore, when calculating processes with twisted photons, the
initial state of the scatterer must be chosen as a wave packet.
Calculations of such processes using the standard S-matrix
formalism, which assumes a description of the scatterer state
using a plane wave and subsequent averaging over the impact
parameter, lead to a loss of dependence of the probability of
processes on the most important characteristic of a Bessel
cylindrical wave---the projection of its angular momentum
\cite{Ivanov,Kiselev}. At the same time, taking into account the
impact parameter and the size of the target, for example, in paper
\cite{Serbo}, where the process of photoionization of a hydrogen
atom by a cylindrical wave was considered in the approximation of
the infinitely heavy nucleus, a dependence of the probability of
the process on both the impact parameter and the magnitude of the
angular momentum of the cylindrical wave was obtained.

In order to understand this in more detail, in this paper we
consider Compton ionization of a hydrogen atom by twisted photons
in the non-relativistic approximation, taking into account the
size of the target and the mass of the nucleus. Earlier, in the
case of photons described by a plane electromagnetic wave, this
process was considered in the papers \cite{HCVP,SVP,SVP1}.  For
brevity, we will further call such photons the plane wave photons.

The work uses the atomic system of units: $m_e=\hbar=|e|=1$. In
these units,  the speed of light $c=137$ and the fine structure
constant $\alpha=1/c$. The classical radius of the electron
$r_0=\alpha^2$, the mass of the proton $M=1836$ a.u. For ease of
perception of the material, we will adhere to the same notations
as in review \cite{Serbo}.

\section{The Hamiltonian}

The basic formulas for obtaining the matrix element with plane
wave photons in the case of the hydrogen atom are presented in
many papers but here it is more convenient for us to adhere to
paper \cite{HCVP}, and we will present only those of them that are
necessary for understanding the presented material. The
non-relativistic Hamiltonian of the hydrogen atom in an
electromagnetic field has the following form (particle 1 is an
electron, particle 2 is a nucleus):
\begin{equation}\label{total_H}
H = \frac{1}{2}\left(-i\vec\nabla_{r_1} - \frac{1}{c}\vec A(\vec
r_1,t)\right)^2+ \frac{1}{2M}\left(-i\vec\nabla_{r_2} +
\frac{1}{c}\vec A(\vec r_2,t)\right)^2-\frac{1}{|\vec r_1-\vec
r_2|}.
\end{equation}
Here $\vec A(\vec r,t)$ is the vector potential of the second
quantized electromagnetic  field in the Coulomb gauge $\vec\nabla
\vec A(\vec r,t)=0$, which we write in the form
\begin{equation}\label{vector_potential_2q}
\frac1c\vec A(\vec r,t) = \sqrt{\frac{2\pi}{\omega}}\int \frac{d^3
k }{(2\pi)^3} \sum_\Lambda \left( e^{i(\vec k \vec r - i\omega t)}
 \vec{e}_{\vec k \Lambda} a_\Lambda(\vec k) +
e^{-i(\vec k \vec r - i\omega t)} \vec{e}^{\,*}_{\vec k \Lambda}
a^\dagger_\Lambda(\vec k)\right),
\end{equation}
where $\vec{e}_{\vec k \Lambda}$, $(\vec k,\vec{e}_{\vec k
\Lambda})=0$, denotes  the polarization vector of a photon with
the momentum $\vec k $, energy $\omega = kc$, and helicity
$\Lambda = \pm 1$, and the photon creation and annihilation
operators are normalized in such a way that the following
relations are satisfied:
\begin{equation}\label{commutator}
\left[a_\Lambda(\vec k), a^\dagger_{\Lambda^\prime}(\vec
k^\prime)\right] = (2\pi)^3 \, \delta_{\Lambda \Lambda^\prime} \, \delta
(\vec k - \vec k^\prime), \quad H_{em} = \int \frac{d^3 k
}{(2\pi)^3}\ \omega(\vec k) \sum_\Lambda a^\dagger_\Lambda(\vec
k) \, a_\Lambda(\vec k).
\end{equation}
Note that in these formulas plane waves are normalized to  one
photon with a certain momentum per unit volume, which turns out to
be convenient when calculating the probabilities of processes, and
the commutation relations are consistent with such normalization.

We consider the initial photon to be twisted but there is no
reason to consider the final photon to be twisted as well. Indeed,
twisted photons are created artificially by passing a plane wave
through a special diffraction grating, and, formally, to register
them, a hypothetical infinitely large detector is needed, capable
of measuring the amplitude and phase of a cylindrical wave at each
point in the plane orthogonal to the direction of its propagation.
Standard detectors in particle physics have finite dimensions and
are adapted to measure particle momenta. Therefore, we will
consider the final photon to be a plane wave one, that is, the
state of the final photon is the standard single-particle state
with a certain momentum,
\begin{equation}\label{photon_f}
| \vec k_1, \Lambda_1 \rangle = a^\dagger_{\Lambda_1}(\vec k_1) \, |0\rangle.
\end{equation}

The initial cylindrical wave can also be conveniently written as a
superposition of plane waves, which will allow us to relate the
matrix element of the photoprocess involving the twisted photon to
a similar matrix element involving a plane wave photon. Assuming
that the cylindrical electromagnetic wave in our problem
propagates along the $z$-axis, we write its vector potential as
(recall that the notation of paper \cite{Serbo} is used):
\begin{equation}\label{vector_potential}
\frac1c\vec A_{\kappa m k_z \Lambda}(\vec r,t) =
\sqrt{\frac{2\pi}{\omega}}\int a_{\kappa m}(\vec k_\perp)\
\vec{e}_{\vec k \Lambda}e^{i(\vec k \vec r - i\omega t)}\
\frac{d^2 k_\perp}{(2\pi)^2} ,
\end{equation}
where
\begin{equation}
a_{\kappa m}(\vec k_\perp)= i^{-m} e^{i m \phi_k}
\frac{2\pi}{k_\perp}\,\delta(k_\perp -\kappa), \qquad  \vec k = (\vec
k_\perp, k_z), \qquad \omega=c\sqrt{\kappa^2+k_z^2},
\end{equation}
and the quantities $\kappa$ and $k_z$, called the transverse and
longitudinal momentum of the twisted photon, are considered fixed.
The polar and azimuthal angles of the momentum of the plane wave
components of the initial twisted photon are denoted by
$(\theta,\phi_k)$. The polar angle of the photon (it is also
called the opening angle) is defined as ${\rm tg}\,\theta=\kappa/k_z$.
Usually, the so-called paraxial waves are considered, for which
$\kappa \ll k_z.$

This choice of vector potential is consistent with the
normalization of plane waves in formula
(\ref{vector_potential_2q})  and differs from that used in
\cite{Serbo} by the factor $ \sqrt{{2\pi}/{\omega}}$. In formula
(\ref{vector_potential}) it is convenient to integrate with
respect to $|\vec k_\perp| = k_\perp$, as a result of which it
takes the form
\begin{equation}\label{vector_potential_1}
\frac1c\vec  A_{\kappa m k_z \Lambda}(\vec r,t) =
\sqrt{\frac{2\pi}{\omega}}\int\limits_0^{2\pi} \, \frac{d \phi_k }{2\pi}\
i^{-m} e^{i m \phi_k}\ \vec{e}_{\vec k(\phi_k) \Lambda}e^{i\left(\vec
k(\phi_k) \vec r - i\omega t\right)},
\end{equation}
where the vector $\vec k(\phi_k)$ is written through the angles $(\theta,\phi_k)$ in the standard way:
\begin{equation}\label{momentum}
\vec k(\phi_k) = k \, (\sin\theta \cos\phi_k, \, \sin\theta \sin
\phi_k, \, \cos\theta).
\end{equation}
Therefore, the state of the initial twisted photon in the second
quantized  theory can be written as an expansion in
single-particle states with a certain momentum as
\begin{equation}\label{photon_ini}
|\kappa, m, k_z, \Lambda \rangle = \int\limits_0^{2\pi} \, \frac{d \phi_k
}{2\pi}\ i^{-m} e^{i m \phi_k}\ |\vec k(\phi_k), \Lambda \rangle.
\end{equation}
It is easy to verify that vector potential (\ref{vector_potential_2q}) of the second quantized electromagnetic field satisfies the relation
\begin{equation}\nonumber
\langle 0|\,\frac1c\vec A(\vec r,t)\,|\kappa, m, k_z, \Lambda\rangle =
\sqrt{\frac{2\pi}{\omega}}\int\limits_0^{2\pi} \, \frac{d \phi_k }{2\pi}\
i^{-m} e^{i m \phi_k}\ \vec{e}_{\vec k(\phi_k) \Lambda}e^{i\left(\vec
k(\phi_k) \vec r - i\omega t\right)} = \frac1c\vec  A_{\kappa m k_z
\Lambda}(\vec r,t).
\end{equation}

\section{Calculation of process probabilities}

To calculate the matrix element of the hydrogen ionization process
in the Born approximation, taking into account the mass of the
nucleus, it is convenient to express the coordinates $\vec r_1$
and $\vec r_2$ through the relative coordinate $\vec{\rho}=\vec
r_1-\vec r_2$ and the coordinate of the center of mass of the atom
$\vec R = (\vec r_1+M\vec r_2)/(M+1)$, where $M=1836$ a.u.:
\begin{equation}
\vec r_1=\vec R+\frac{M}{M+1}\vec\rho, \qquad \vec r_2=\vec R
-\frac{1}{M+1} \vec\rho.
\end{equation}
Then for the conjugate momenta we have the expressions
\begin{equation} \label{momenta}
\vec p_e=\frac{1}{M+1}\vec P_R+\vec p_\rho, \qquad \vec
P=\frac{M}{M+1}\vec P_R-\vec p_\rho.
\end{equation}
Here $\vec p_e$ is the electron momentum, $\vec P$ is the proton
(nucleus) momentum, $\vec P_R$ is the total momentum, $\vec
p_\rho$ is the relative momentum of the fragments.

For the plane wave photons, the conservation of momentum and
energy in the laboratory frame takes the form (it is assumed that
the atom is stationary before the collision):
\begin{equation}
\vec k = \vec k_1+\vec p_e + \vec P , \qquad \omega+\varepsilon_0 = \frac12 p^2_e +\omega_1 + \frac{1}{2M}P^2,
\end{equation}
where $\vec k$ and $\vec k_1$ ($\omega$ and $\omega_1$) are the
momenta (energies)  of the photon before and after the
interaction, respectively, and $\varepsilon_0= -1/2$ is the energy
of the ground state of the hydrogen atom. Denoting the transferred
momentum as
\begin{equation} \label{Q}
\vec Q=\vec k-\vec k_1,
\end{equation}
from this we get $\vec Q=\vec P_R$.

The states of the system of a proton and an electron interacting
with  a second quantized electromagnetic field are described by
tensor products of the wave functions of the proton and electron
system and vectors in the Hilbert space of states of the second
quantized electromagnetic field. We will work in the $A^2$
approximation, that is, we will take into account only the terms
quadratic in the electromagnetic field in  Hamiltonian
(\ref{total_H}). When calculating the matrix element, it is
convenient to calculate first the matrix element between the
states of the electromagnetic field, which results in an effective
potential of interaction between the proton and electron, and then
calculate the matrix element between the states of the proton and
electron system. Thus, we need to calculate
\begin{equation}\label{matrix_el}
{\mathcal M}^{tw} = \int\limits_{-\infty}^{\infty}dt \, \langle\Psi^-(\vec
p_e,\vec P,t)| \, {\tilde V}_{int}(t) \, |\Phi_0(t) \rangle,
\end{equation}
where the effective interaction potential is written as
\begin{equation}
\begin{split}
& {\tilde V_{int}}(\vec R, \vec\rho; t) = \langle\vec k_1, \Lambda_1|
\ \frac{1}{2}\left( \frac{1}{c}\vec A(\vec r_1,t)\right)^2+
\frac{1}{2M}\left( \frac{1}{c}\vec A(\vec r_2,t)\right)^2 |\kappa,
m, k_z, \Lambda \rangle  =  \\
& = \frac{\pi}{\sqrt{\omega\omega_1}}\int \,\frac{d \phi_k }{2\pi}\
i^{-m} e^{i m \phi_k}\, \big(\vec e_{\vec k(\phi_k) \Lambda},\vec
e^{\,*}_{\vec k_1 \Lambda_1}\big)\, e^{i\left(\vec Q\vec R-(\omega-\omega_1)
t\right)}  \left( e^{i\vec Q\vec\rho} + \frac{1}{M}e^{-i(1/M)\vec
Q\vec\rho}\right).
\end{split}
\end{equation}

We take the wave function of the final state in the Born approximation in the form
\begin{equation}\label{WFF}
\langle \vec r_1,\vec r_2 | \Psi^-(\vec p_e,\vec P,t) \rangle = e^{i\vec P_R\vec R
- i(E_1+E_2) t} \, \phi^-(\vec p_\rho,\vec\rho),
\end{equation}
where $E_1 = p_e^2 / 2$, $E_2 = P^2 / 2M$, and the function
$\phi^-(\vec p_\rho,\vec\rho)$ satisfies the Schr\"odinger
equation for the hydrogen atom in the relative coordinates and
describes the state of the continuous spectrum,
\begin{equation}\label{Schreq}
\left(\frac{1}{2\mu} p^2_\rho +\frac{1}{2\mu}
\triangle_{\rho}+\frac{1}{\rho}\right)\phi^-(\vec
p_\rho,\vec\rho)=0, \qquad \mu=\frac{M}{M+1}\approx 1.
\end{equation}

The problem of choosing the initial state wave function in the
case of processes with twisted photons turns out to be more
complicated. The field of twisted photons
(\ref{vector_potential_1}) is not translationally invariant in the
directions orthogonal to the $z$-axis, so the wave function of the
center of mass motion cannot be taken in the commonly used form of
a plane wave. In the case of twisted photons, this must be a wave
function corresponding to a localized state. Therefore, we choose
the initial wave function of the hydrogen atom, for example, as
the product of the wave function of the ground state of the
hydrogen atom $\phi_0(\rho)$, which depends on the relative
coordinate $\vec{\rho}$ and is a solution of equation
(\ref{Schreq}), and the wave function of the center of mass
motion, which depends on $\vec R$ and is described by a Gaussian
wave packet centered at point $\vec b$. Without loss of
generality, we can assume that the third coordinate of the vector
$\vec b$ is equal to zero, that is, it is actually an impact
parameter. If this state has a macroscopic size, for example, if
$d \sim 0.1$ cm, then such a state allows us to phenomenologically
take into account the size and location of a macroscopic target.
The average energy of the state described by such a wave packet is
$\langle \frac{\vec P_R^2}{2M} \rangle \sim 10^{-17}$ eV, and the
lifetime of this state is $\tau = \frac{M d^2}{\hbar} \sim 1$ sec,
that is, macroscopic. Formally, when calculating a matrix element
with such a state, we would have to integrate over a time interval
of the order of the lifetime of this state. In this case, the
energy would be conserved with an accuracy of ${\hbar}/ \tau =
{1}/{M d^2} \sim 10^{-17}$ eV, which, of course, cannot be
observed experimentally. Therefore, the use of such a state to
describe the process of Compton ionization of a hydrogen atom in a
field of twisted photons is quite correct. However, to simplify
the calculations, we will just neglect the dependence of this
state on time, that is, we will take the wave function of the
initial state in the form
\begin{equation}\label{WFI}
\langle \vec r_1,\vec r_2 | \Phi_0(t) \rangle = \frac{1}{(\sqrt{\pi}d)^{3/2}}\
e^{-i\varepsilon_0 t}\, \phi_0(\rho)\, e^{-\frac{(\vec R - \vec
b)^2}{2 d^2}}
\end{equation}
and when calculating the matrix element we will integrate with
respect to time within infinite limits.

It should be emphasized that  a correct design of an experiment on
Compton (or conventional) photoionization of atoms by twisted
photons discussed in this paper implies  that the condition
$\kappa d \gg 1$ is fulfilled. Indeed, in such processes, the
final state contains an electron in a continuous spectrum state,
and therefore it is necessary to take into account the momentum
conservation law. Due to the uncertainty principle, for a target of
size $d$, the momentum is conserved with an accuracy of $1/d$. If
in an experiment the condition $\kappa d \sim 1$ appears to be
valid, then the transverse momentum $\kappa$ turns out to be of
the order of the uncertainty in determining the momentum, which,
in this case, will not allow to measure the characteristics of the
scattering process with an accuracy sufficient  to detect the
effect of the cylindrical wave.

Substituting wave functions (\ref{WFF}) and (\ref{WFI}) into formula
(\ref{matrix_el}) and integrating with respect to time, we obtain:
\begin{equation}\label{matrix_el1}
\begin{gathered}
{\mathcal M}^{tw}_{\Lambda\Lambda_1}\big(\vec k_1, \vec p_e, \vec P\big)=
i^{-m} \frac{2\pi^2}{\sqrt{\omega\omega_1}} \, \delta(\omega
-|\varepsilon_0|-\omega_1-E_1-E_2 )\, \times \\
\times \int \frac{d \phi_k }{2\pi}\, e^{i m \phi_k}\, \big(\vec
e_{\vec k(\phi_k) \Lambda},\vec e^{\,*}_{\vec k_1 \Lambda_1}\big)\,
{\mathcal M}^{pl}\big(\vec k(\phi_k), \vec k_1\ \vec p_e, \vec P\big) \, \times \\
\times \int d^3 R\ e^{i\vec R \left(\vec k(\phi_k)-\vec
k_1-\vec p_e-\vec P\right)}\,
\frac{1}{(\sqrt{\pi}d)^{3/2}} \, e^{-\frac{(\vec R - \vec b)^2}{2
d^2}} .
\end{gathered}
\end{equation}
Here the indices $tw$ and $pl$ denote the matrix elements
calculated with twisted and plane wave initial photons, the latter
being written as
\begin{equation} \label{M_pl}
{\mathcal M}^{pl}\big(\vec k(\phi_k), \vec k_1, \vec p_e, \vec P\big) = \langle \phi^-(\vec
p_\rho)|\left( e^{i\vec Q\vec\rho} + \frac{1}{M}e^{-i(1/M)\vec
Q\vec\rho}\right)|\phi_0 \rangle.
\end{equation}
Integrating with respect to $\vec R$ and temporarily omitting the
energy  conservation delta function multiplied by $2\pi$, we
rewrite formula (\ref{matrix_el1}) as
\begin{equation} \label{M_tw}
\begin{gathered}
{\mathcal M}^{tw}_{\Lambda\Lambda_1}\big(\vec k_1, \vec p_e, \vec P\big)=
i^{-m} \, \frac{\pi (2\sqrt{\pi}d)^{3/2}}{\sqrt{\omega\omega_1}}
 \int \frac{d\phi_k }{2\pi} \, e^{i m \phi_k}\, e^{i\vec b \left(\vec k(\phi_k)-\vec k_1-\vec p_e-\vec P\right)}
 \times \\
 \times \, e^{-\frac{d^2}{2} \left(\vec k(\phi_k)-\vec k_1-\vec
p_e-\vec P\right)^2} \, \big(\vec e_{\vec k(\phi_k) \Lambda},\vec e^{\,*}_{\vec
k_1 \Lambda_1}\big)\, {\mathcal M}^{pl}\big(\vec k(\phi_k), \vec k_1, \vec p_e, \vec
P\big).
\end{gathered}
\end{equation}
Then the squared matrix element, in which the summation over the polarizations of the final photon is performed, is written as
\begin{equation} \label{matrix_el^2}
\begin{gathered}
\sum\limits_{\Lambda_1} \big|{\mathcal M}^{tw}_{\Lambda\Lambda_1}\big(\vec k_1, \vec p_e, \vec
P\big)\big|^2 =  \frac{\pi^2
(2\sqrt{\pi}d)^{3}}{\omega\omega_1}\int\limits_0^{2\pi}\frac{d\phi_k}{2\pi}\
\int\limits_0^{2\pi}\frac{d\phi^\prime_k}{2\pi}\, \times
\\
\times \, e^{i m(\phi_k -\phi^\prime_k)}\, e^{i\left(\vec
k_\perp(\phi_k) - \vec k_\perp(\phi^\prime_k)\right) \vec b}\,
e^{-\frac{d^2}{2} \left(\vec k(\phi_k)-\vec k_1-\vec p_e-\vec P\right)^2} \,
e^{-\frac{d^2}{2} \left(\vec k(\phi^\prime_k)-\vec k_1-\vec p_e-\vec
P\right)^2}\,  \times
\\
\times \left(\big(\vec e_{\vec k(\phi_k) \Lambda},\vec e^{\,*}_{
\vec k(\phi^\prime_k) \Lambda}\big) - \frac{\big(\vec e_{\vec
k(\phi_k)\Lambda}, \vec k_1\big)\big(\vec e^{\,*}_{\vec k(\phi^\prime_k)
\Lambda}, \vec k_1\big)}{\vec k_1^2}\right) \times
\\
\times \, {\mathcal M}^{pl}\big(\vec k(\phi_k), \vec k_1, \vec p_e,
\vec P\big) \, {\mathcal M}^{pl*}\big(\vec k(\phi^\prime_k), \vec k_1, \vec p_e, \vec
P\big),
\end{gathered}
\end{equation}
where we used the standard formula for the summation over the photon polarizations,
\begin{equation}
\sum_{\Lambda_1} \big(\vec e_{\vec k \Lambda},\vec e^{\,*}_{\vec k_1
\Lambda_1}\big) \big(\vec e^{\,*}_{\vec k^\prime \Lambda^\prime},\vec
e_{\vec k_1 \Lambda_1}\big) = \big(\vec e_{\vec k \Lambda},\vec
e^{\,*}_{\vec k^\prime \Lambda^\prime}\big) - \frac{\big(\vec e_{\vec k
\Lambda},\vec k_1\big)\big(\vec e^{\,*}_{\vec k^\prime
\Lambda^\prime},\vec k_1\big)}{\vec k_1^2}.
\end{equation}

If the target size $d$ is macroscopic, for example, of the order
of $0.1$ cm,  then the exponentials containing $d^2$ are
practically equal to zero when the sum of the momenta in the
exponent differs from zero by more than $10^{-3}$ eV. This can be
taken into account as follows. Let us pass  to the new variables
$\phi_+$ and $\phi_-$ in the double integral in formula
(\ref{matrix_el^2}) using the formulas
\begin{equation}\nonumber
\phi_\pm = \frac{\phi_{k} \pm \phi^\prime_{k}}{2}, \quad 0 \leq
\phi_+ < 2\pi, \quad -\pi \leq \phi_- < \pi, \quad \frac{
D(\phi_{k}, \phi^\prime_{k})}{ D(\phi_-, \phi_+)} = 2.
\end{equation}
Then formula (\ref{matrix_el^2}) will take the form
\begin{equation}\label{matrix_el^2m}
\begin{gathered}
\sum\limits_{\Lambda_1} \big|{\mathcal
M}^{tw}_{\Lambda\Lambda_1}\big(\vec k_1, \vec p_e, \vec
P\big)\big|^2 = \frac{2\pi^2
(2\sqrt{\pi}d)^{3}}{\omega\omega_1}\int\limits_0^{2\pi}
\frac{d\phi_+}{2\pi} \int\limits_{-\pi + |\phi_+ - \pi|}^{\pi
-|\phi_+ - \pi|}\frac{d\phi_-}{2\pi}\, \times
\\
\times \, e^{2i m\phi_-}\, e^{i\left(\vec k_\perp(\phi_k) - \vec
k_\perp(\phi^\prime_k)\right) \vec b}\, e^{-\frac{d^2}{2} \left[\left(\vec k(\phi_+ + \phi_-)-\vec k_1-\vec p_e-\vec
P\right)^2 + \left(\vec k(\phi _+ - \phi_-)-\vec k_1-\vec p_e-\vec P\right)^2\right]}\, \times
\\
\times \left(\big(\vec e_{\vec k(\phi_k) \Lambda},\vec
e^{\,*}_{\vec k(\phi^\prime_k) \Lambda}\big) - \frac{\big(\vec
e_{\vec k(\phi_k)\Lambda},\vec k_1\big)\big(\vec e^{\,*}_{\vec
k(\phi^\prime_k) \Lambda},\vec k_1\big)}{\vec k_1^2}\right) \times
\\
\times \, {\mathcal M}^{pl}\big(\vec k(\phi_k), \vec k_1, \vec
p_e, \vec P\big) \, {\mathcal M}^{pl*}\big(\vec k(\phi^\prime_k),
\vec k_1, \vec p_e, \vec P\big) .
\end{gathered}
\end{equation}
Here it is implied that $\phi_k = \phi_+ + \phi_-$, $\phi^\prime_k
= \phi_+ - \phi_-$,  where this substitution is not explicitly
written out.

We identically transform the terms in the square brackets in the
exponent containing $d^2$ as follows:
\begin{equation}
\begin{gathered}
\frac{d^2}{2} \left[\big(\vec k(\phi_+ + \phi_-)-\vec k_1-\vec p_e-\vec
P\big)^2 + \big(\vec k(\phi _+ - \phi_-)-\vec k_1-\vec p_e-\vec P\big)^2\right]
\equiv \\
\equiv \frac{d^2}{4} \big(\vec k(\phi_+ + \phi_-) - \vec k(\phi_+ -
\phi_-)\big)^2 + {d^2} \left[\frac{1}{2} \big(\vec k(\phi_+ + \phi_-) + \vec
k(\phi_+ - \phi_-)\big) - \vec k_1 - \vec p_e-\vec P\right]^2,
\end{gathered}
\end{equation}
and using formula (\ref{momentum}) we write the expression in the
first bracket in the r.h.s. of this formula in terms of the
angles:
\begin{equation}
\frac{ d^2}{4} \big(\vec k(\phi_+ + \phi_-) - \vec k(\phi_+ - \phi_-)\big)^2 =\kappa^2 d^2 \sin^2\phi_-.
\end{equation}
Substituting these expressions into formula (\ref{matrix_el^2m}), we rewrite it in the form
\begin{equation}\label{matrix_el^2m2}
\begin{gathered}
\sum\limits_{\Lambda_1} \big|{\mathcal
M}^{tw}_{\Lambda\Lambda_1}\big(\vec k_1, \vec p_e, \vec
P\big)\big|^2 = \frac{2\pi^2
(2\sqrt{\pi}d)^{3}}{\omega\omega_1}\int\limits_0^{2\pi}
\frac{d\phi_+}{2\pi} \int\limits_{-\pi + |\pi - \phi_+ |}^{\pi
- |\pi - \phi_+ |}\frac{d\phi_-}{2\pi}\, \times \\
\times \, e^{2i m\phi_-}\, e^{i\left(\vec k_\perp(\phi_k) - \vec
k_\perp(\phi^\prime_k)\right) \vec b}\, e^{-\kappa^2 d^2
\sin^2{\phi_-}}\,
 e^{- d^2\left[\frac{1}{2} \left(\vec k(\phi_+ + \phi_-) + \vec k(\phi _+ -
 \phi_-)\right) - \vec k_1 - \vec p_e-\vec P\right]^2} \, \times \\
\times \left(\big(\vec e_{\vec k(\phi_k) \Lambda},\vec
e^{\,*}_{\vec k(\phi^\prime_k) \Lambda}\big) - \frac{\big(\vec
e_{\vec k(\phi_k)\Lambda},\vec k_1\big)\big(\vec e^{\,*}_{\vec
k(\phi^\prime_k) \Lambda},\vec k_1\big)}{\vec k_1^2}\right) \times \\
\times \, {\mathcal M}^{pl}\big(\vec k(\phi_k), \vec k_1, \vec p_e,
\vec P\big)\,{\mathcal M}^{pl*}\big(\vec k(\phi^\prime_k), \vec k_1, \vec p_e, \vec P\big) .
\end{gathered}
\end{equation}

For $\kappa d \gg 1$, the presence of the factor $e^{-\kappa^2 d^2
\sin^2{\phi_-}}$  in the integrand leads to the fact that the main
contribution to the integral with respect to $ \phi_-$ comes from
a very narrow region near $ \phi_- = 0$, which is present for any
value of $ \phi_+$, and very narrow regions near $ \phi_- =
\pm\pi$, which are present only for the value of $ \phi_+$ in a
very narrow region near $ \phi_+ = \pi$. Therefore, the
contribution of the regions near $ \phi_- = \pm\pi$ can be
neglected and the integration with respect to $\phi_-$ can be
limited to the interval $-\frac{\pi}{2} \leq \phi_- <
\frac{\pi}{2}$ for any value of $ \phi_+$.

Assuming that the matrix elements ${\mathcal M}^{pl}$, the photon
polarization vectors,  and the exponential factor containing the
electron and nucleus momenta smoothly depend on the angle $
\phi_-$, we can pull them out of the integral with respect to $
\phi_-$, putting $ \phi_- = 0$ in them. In this case, $\phi_k =
\phi^\prime_k = \phi_+ $, and the sum over the polarizations is
transformed as follows:
\begin{equation}
\left(\big(\vec e_{\vec k(\phi_+) \Lambda},\vec e^{\,*}_{\vec
k(\phi_+) \Lambda}\big) - \frac{\big(\vec e_{\vec k(\phi_+)\Lambda},\vec
k_1\big)\big(\vec e^{\,*}_{\vec k(\phi_+) \Lambda},\vec k_1\big)}{\vec
k_1^2}\right) = \frac{1}{2}\left(1 + \frac{\big(\vec k(\phi_+) , \vec
k_1\big)^2}{\vec k^2(\phi_+)\,\vec k_1^2}\right).
\end{equation}
As a result, formula (\ref{matrix_el^2m2}) takes the form
\begin{equation}\label{matrix_el^2m3}
\begin{gathered}
\sum\limits_{\Lambda_1} \big|{\mathcal
M}^{tw}_{\Lambda\Lambda_1}\big(\vec k_1, \vec p_e, \vec
P\big)\big|^2 = \frac{\pi^2
(2\sqrt{\pi}d)^{3}}{\omega\omega_1}\int\limits_0^{2\pi}
\frac{d\phi_+}{2\pi} \left(1 + \frac{\big(\vec k(\phi_+),
\vec k_1\big)^2}{\vec k^2(\phi_+)\,\vec k_1^2}\right)
e^{- d^2\left( \vec k(\phi_+ ) - \vec k_1 - \vec p_e-\vec P\right)^2} \times \\
\times \, \big|{\mathcal M}^{pl}\big(\vec k(\phi_+), \vec k_1, \vec p_e, \vec P\big)\big|^2
\int\limits_{-\frac{\pi}{2}}^{\frac{\pi}{2}}\frac{d\phi_-}{2\pi}\,
  e^{2i m\phi_-}\, e^{i\left(\vec k_\perp(\phi_+ + \phi_-) - \vec k_\perp(\phi_+ - \phi_-)\right)
\vec b}\, e^{-\kappa^2 d^2 \sin^2\phi_-}.
\end{gathered}
\end{equation}
Now we need to evaluate the integral
\begin{equation}\label{int_0}
I_{\phi_-} \equiv  \int\limits_{-\frac{\pi}{2}}^{\frac{\pi}{2}}\frac{d\phi_-}{2\pi}\,
  e^{2i m\phi_-}\, e^{i\left(\vec k_\perp(\phi_+ + \phi_-) - \vec k_\perp(\phi_+ - \phi_-)\right)
\vec b}\, e^{-\kappa^2 d^2 \sin^2\phi_-}.
\end{equation}
Without loss of generality we can assume that $\vec b = (b, 0,0).$ In this case
\begin{equation*}
\big(\vec k_\perp(\phi_+ + \phi_-) - \vec k_\perp(\phi_+ - \phi_-)\big)
\vec b = \kappa b \big(\cos(\phi_+ + \phi_-) - \cos(\phi_+ - \phi_-)\big)
= -2 \kappa b\sin\phi_+ \sin\phi_-,
\end{equation*}
and integral \eqref{int_0} takes the form
\begin{equation}
I_{\phi_-} =  \int\limits_{-\frac{\pi}{2}}^{\frac{\pi}{2}}\frac{d\phi_-}{2\pi}\,
  e^{2i m\phi_-}\, e^{-2i \kappa b\sin\phi_+ \sin\phi_-} \, e^{-\kappa^2 d^2
  \sin^2\phi_-}.
\end{equation}
Since the main contribution to the last integral is given by the
values of $\phi_-$ close to zero, we replace $\sin\phi_-$
everywhere in the integrand with $\phi_-$, pass to the variable $x
= \kappa d \phi_-$ and extend the integration with respect to this
variable to $-\infty$ and $\infty$. As a result, we obtain:
\begin{equation}\label{integral1}
I_{\phi_-} \simeq \frac{1}{\kappa d}
\int\limits_{-\infty}^{\infty}\frac{dx}{2\pi}\,
  e^{\frac{2i m x}{\kappa d}}\, e^{-\frac{2i b\sin\phi_+ x}{d}} \, e^{-x^2} =
  \frac{1}{2\sqrt{\pi}\kappa d}\, e^{-\frac{(m -
\kappa b\sin \phi_+)^2 }{\kappa^2 d^2}} .
\end{equation}
Substituting this expression into formula (\ref{matrix_el^2m2}), we finally find
\begin{equation} \label{final_sum_abs_sqr}
\begin{gathered}
\sum_{\Lambda_1} \big|{\mathcal M}^{tw}_{\Lambda\Lambda_1}\big(\vec k_1, \vec p_e, \vec
P\big)\big|^2 =  \frac{ 4 \pi^3
d^2}{\kappa\omega\omega_1}\int\limits_0^{2\pi}\frac{d\phi_+}{2\pi}\,
e^{-\frac{(m - \kappa b\sin \phi_+)^2 }{\kappa^2 d^2}} \left(1 +
\frac{\big(\vec k(\phi_+), \vec k_1\big)^2}{\vec k^2(\phi_+)\,\vec
k_1^2}\right) \times \\
\times \, e^{-{d^2} \left(\vec k(\phi _+ ) - \vec k_1 - \vec p_e-\vec P\right)^2}\,
\big|{\mathcal M}^{pl}\big(\vec k(\phi_+), \vec k_1, \vec p_e, \vec P\big)\big|^2 .
\end{gathered}
\end{equation}

Before moving on, let us check the correctness of the
transformations performed. Since  we made several different
approximations in original formula (\ref{matrix_el^2}), it is
quite difficult to estimate their accuracy analytically. The
simplest way to estimate the accuracy of the approximations made
is to perform the integration in  formula
\eqref{final_sum_abs_sqr} numerically for some realistic values of
the parameters, and then to compare the result with the one
obtained directly from  original expression \eqref{M_tw} after
squaring its modulus and summing over the polarizations
$\Lambda_1$. We choose the target size $d \sim 0.1 \textrm{ cm} =
5 \cdot 10^3 \textrm{ eV}^{-1}$. We immediately note that for the
impact parameter value $b \gg d$ or if $| \vec k(\phi _+ ) - \vec
k_1 - \vec p_e-\vec P | > 10^{-3}$ eV for all $\phi_+$, both
results are practically zero. Therefore, we consider $b \lesssim
d$, as well as such momenta of the particles that there exists an
angle $\phi_+$ for which the approximate conservation law $| \vec
k(\phi _+ ) - \vec k_1 - \vec p_e-\vec P | < 10^{-3}$ eV is valid.

When calculating the plane wave amplitude ${\mathcal M}^{pl}$
using  formula \eqref{M_pl}, for simplicity we neglect the second
term in the matrix element $e^{-i(1/M)\vec Q\vec\rho} / M$, and
take the wave function of the final state $\phi^-(\vec p_\rho)$ in
the form of a plane wave, $ \phi ^ - (\vec p_\rho ) = e^{i\vec
p_\rho \vec \rho }.$ Substituting this into  formula \eqref{M_pl},
we find
\begin{equation}\nonumber
{\mathcal M}^{pl}\big(\vec k(\phi_k), \vec k_1, \vec p_e, \vec
P\big) =\frac{8 \sqrt{\pi} \alpha^\frac{5}{2}}{[(\vec Q-\vec
p_\rho)^2+\alpha^2]^2},
\end{equation}
and taking into account that, in accordance with
\eqref{momenta}--\eqref{Q},
\begin{equation}\nonumber
\vec p_e = \frac{1}{{M + 1}}\vec P_R + \vec p_\rho \approx \vec
p_\rho \quad \Rightarrow \quad \vec Q - \vec p_\rho \approx \vec
P,
\end{equation}
we obtain that, due to the simplifications made, the amplitude
${\mathcal M}^{pl}$ depends only on the free variable $\vec P$ and
can be pulled out of the integral in expressions \eqref{M_tw},
\eqref{final_sum_abs_sqr}.

Taking into account the comments made, we carried out calculations
for several values of each of the parameters, where the remaining
parameters have the following orders of magnitude: $k_z \sim 10$
keV, $\kappa \sim 100$ eV, $m \sim 10^3$, $p_e \sim \omega_1 \sim
5$ keV, and $P$ is taken ``by the residual principle'', such as
to satisfy the law of conservation of momentum with a deviation of
no more than $10^{-3}$ eV. The number of integration lattice nodes
both in the region of the narrow peak determined by the
exponential $e^{-{d^2} \left(\vec k(\phi _+ ) - \vec k_1 - \vec
p_e-\vec P\right)^2}$ and in the rest of the integration region
outside the peak was taken to be equal to $10^3$--$10^6$. The
calculations showed that for such values of the parameters, both
results (from formula \eqref{M_tw}  and formula
\eqref{final_sum_abs_sqr}) coincide with a relative accuracy of at
least $10^{-6}$, which almost does not depend on the size of a
sufficiently large integration lattice. Thus, the numerical study
indicates that the approximations made are sufficiently accurate,
and the calculation methods used are mathematically correct.

Let us return to formula \eqref{final_sum_abs_sqr}. As already
mentioned, the exponential $e^{-{d^2} \left(\vec k(\phi_+ ) - \vec
k_1 - \vec p_e-\vec P\right)^2}$ present there, multiplied by
$d^3/\pi^{3/2}$, behaves in fact as the delta function of the sum
of momenta in its exponent. Therefore, to find the differential
probability of the process of Compton ionization of the hydrogen
atom, we need to integrate this exponential with respect to one of
the momenta and express this momentum in the remaining expression
in terms of the other three momenta in accordance with the
conservation law. We integrate this exponential with respect to
the momentum $\vec P$ of the nucleus,
\begin{equation}
\int e^{-{d^2} \left(\vec k(\phi _+ ) - \vec k_1 - \vec p_e-\vec P\right)^2}
\frac{d^3 P}{(2\pi)^3} = \frac{\sqrt{\pi}}{8\pi^2 d^3},
\end{equation}
and put $\vec P = \vec k(\phi _+ ) - \vec k_1 - \vec p_e \equiv \vec P(\phi _+ )$ in the matrix element.

Taking this formula into account, the differential probability  of
the process of Compton ionization of the hydrogen atom per unit
time, summed over the polarizations of the final photon, is
written as follows:
\begin{equation}\label{dif_probab}
\begin{gathered}
dW =  \frac{\pi^2 \sqrt{\pi}}{\omega\omega_1 \kappa d}
\int\limits_0^{2\pi}\frac{d\phi_+}{2\pi}\, e^{-\frac{(m - \kappa b\sin
\phi_+)^2 }{\kappa^2 d^2}} \, \delta\big(\omega -|\varepsilon_0| -
\omega_1 - \vec p_e^{\,2}/2 -\vec P^2(\phi_+)/2M\big) \, \times \\
\times \left(1 + \frac{\big(\vec k(\phi_+) , \vec
k_1\big)^2}{\vec k^2(\phi_+)\,\vec k_1^2}\right) \big| {\mathcal M}^{pl}\big(\vec
k(\phi_+), \vec k_1, \vec p_e, \vec P(\phi_+)\big)\big|^2 \,
 \frac{d^3 p_e}{(2\pi)^3} \frac{d^3 k_1}{(2\pi)^3}.
\end{gathered}
\end{equation}
Let us recall that, if the recoil momentum of the nucleus is taken
into account, the differential cross section of Compton ionization
of the hydrogen atom by the plane wave photon with momentum $\vec
k(\phi_+)$ can be written in the form \cite{HCVP}
\begin{equation}
\begin{gathered}
d\sigma(\phi_+) =
\frac{(2\pi)^3\alpha}{\omega\omega_1}\,\delta\big(\omega
-|\varepsilon_0| - \omega_1 - \vec p_e^{\,2}/2 - \vec
P^2(\phi_+)/2M\big)\left(1 + \frac{\big(\vec k(\phi_+), \vec
k_1\big)^2}{\vec k^2(\phi_+)\,\vec k_1^2}\right) \times \\
\times \, \big| {\mathcal M}^{pl}\big(\vec k(\phi_+), \vec k_1, \vec p_e, \vec
P(\phi_+)\big)\big|^2 \, \frac{d^3 p_e}{(2\pi)^3} \frac{d^3 k_1}{(2\pi)^3}.
\end{gathered}
\end{equation}
This expression for the cross-section allows us to rewrite formula
(\ref{dif_probab}) as follows:
\begin{equation}\label{dif_probab_1}
dW =\frac{c}{8\sqrt{\pi}\kappa d}
\int\limits_0^{2\pi}\frac{d\phi_+}{2\pi}\, e^{-\frac{(m - \kappa b\sin
\phi_+)^2 }{\kappa^2 d^2}} \, d\sigma(\phi_+),
\end{equation}
where we also took into account that $\alpha=1/c$ in the system of units used. In this formula, the factor
\begin{equation}\nonumber
\frac{c}{{16\pi}\sqrt{\pi}\kappa d} \, e^{-\frac{(m - \kappa b\sin
\phi_+)^2 }{\kappa^2 d^2}}
\end{equation}
gives the flux of photons with momentum $\vec k(\phi_+)$ through the target, while the integral
\begin{equation}
\frac{c}{{16\pi}\sqrt{\pi}\kappa d} \int\limits_0^{2\pi}{d\phi_+}\,
e^{-\frac{(m - \kappa b\sin \phi_+)^2 }{\kappa^2 d^2}}
 = \frac{c}{{16\pi}\sqrt{\pi}\kappa d}\, I
\end{equation}
gives the total flux of photons of the cylindrical wave through
the target.  Therefore, for a specific cylindrical wave and a
specific target, one can define a generalized differential
scattering cross section as the ratio of the probability of the
process to the flux,
\begin{equation}\label{cross-section}
d\bar \sigma = \frac{1}{I} \int\limits_0^{2\pi}{d\phi_k}\, e^{-\frac{(m -
\kappa b\sin \phi_k)^2 }{\kappa^2 d^2}} \, d\sigma(\phi_k) =
d\sigma(\phi^*_k),
\end{equation}
where $\phi^*_k$ is some fixed angle, which is determined by  the
parameters of the cylindrical wave and the target and, in turn,
determines the momentum $\vec k^* = \vec k(\phi^*_k)$  by formula
(\ref{momentum}). The last equality in formula
(\ref{cross-section}) is a consequence of the mean value theorem
for definite integrals.

Note that formula (\ref{dif_probab_1}) for the probability of the
Compton ionization process indicates that for twisted photons,
unlike plane wave photons, there is no natural universal
definition of the differential scattering cross section applicable
to all cylindrical waves and all targets. The generalized
differential cross section (\ref{cross-section}) depends on both
the characteristics of the cylindrical wave and the
characteristics of the target. This is due to the fact that, for
twisted photons, the flux cannot be defined universally, because
it varies depending on the point in the coordinate or momentum
space. The latter is clearly seen in the formulas found for the
probability of the process.

Moreover, formula (\ref{cross-section}) shows that the
differential  probability calculated for a specific target with a
cylindrical wave always coincides with the differential
probability calculated for a plane electromagnetic wave with some
momentum $\vec k^*$, i.e. the use of cylindrical waves does not
give new angular distributions. It is easy to see that this
conclusion is also valid for the processes of photoionization by
twisted photons, which were considered in papers \cite{Serbo,
Kiselev}.

\section{Conclusion}
In the present paper, a method for calculating the differential
probability of the  process of Compton ionization of a hydrogen
atom by a cylindrical electromagnetic wave is developed, which
allows taking into account the finite dimensions of the target and
obtaining the dependence of this probability on the projection of
the angular momentum of the cylindrical wave. The generalized
differential cross section is also defined for this process, which
is obtained by averaging the usual differential cross section for
a plane wave over the transverse momenta of the cylindrical wave
and coincides with the usual cross section for a certain value of
the transverse momentum, which depends both on the characteristics
of this wave and on the characteristics of the target. The latter
means that the differential probability of such a process
coincides with the differential probability of Compton ionization
of a hydrogen atom by a certain plane electromagnetic wave. The
same situation takes place for the usual photoionization
processes. Thus, in the case of  finite-size targets, the use of
cylindrical waves does not lead to new angular distributions, i.e.
experiments on the ionization of atoms by twisted photons do not
actually allow obtaining new information about the structure of
the target atoms.

\section*{Acknowledgments}
The authors are grateful to Yu.V. Popov, V.G. Serbo and
M.N.~Smolyakov for numerous useful discussions.

\end{document}